\documentstyle[12pt]{article}
\def\ref{\par\hangindent=1.0cm\hangafter=1}

\begin{document}

\baselineskip=21pt plus 1pt minus 1pt

\centerline{\bf GENERALIZED DEFORMED su(2) ALGEBRAS,}

\centerline{\bf  DEFORMED PARAFERMIONIC OSCILLATORS AND FINITE W ALGEBRAS}

\bigskip\bigskip
\centerline{DENNIS BONATSOS $^{*+}$\footnote[1]{e-mail:
bonat@ectstar.ect.unitn.it, bonat@cyclades.nrcps.ariadne-t.gr},
C. DASKALOYANNIS
$^{\#}$\footnote[2]{e-mail: daskaloyanni@olymp.ccf.auth.gr},
and P. KOLOKOTRONIS $^+$}

\centerline{$^*$  European Centre for Theoretical Studies in Nuclear
Physics and Related Areas (ECT$^*$)}

\centerline{Strada delle Tabarelle 286, I-38050 Villazzano (Trento), Italy}

\centerline{$^+$ Institute of Nuclear Physics, NCSR ``Demokritos''}

\centerline{GR-15310 Aghia Paraskevi, Attiki, Greece}

\centerline{$^{\#}$ Department of Physics, Aristotle University of
Thessaloniki}

\centerline{GR-54006 Thessaloniki, Greece}

\bigskip\bigskip\bigskip

\centerline{\bf Abstract}

Several physical systems (two identical particles in two dimensions,
iso\-tro\-pic oscillator and Kepler system in a 2-dim curved space) and
mathematical structures (quadratic algebra QH(3), finite W algebra
$\bar {\rm W}_0$) are shown to posses the structure of a generalized
deformed su(2) algebra, the representation theory of which is known.
Furthermore, the generalized deformed parafermionic oscillator is
identified with the algebra of several physical systems (iso\-tro\-pic
oscillator and Kepler system in 2-dim curved space, Fokas--Lagerstrom,
Smorodinsky--Winternitz and Holt potentials)
and mathematical constructions (generalized deformed su(2) algebra,
finite W algebras $\bar {\rm W}_0$ and W$_3^{(2)}$). The fact that the
Holt potential is characterized by the W$_3^{(2)}$ symmetry is obtained
as a by-product.

\vfill\eject

{\bf 1. Introduction}

Quantum algebras (quantum groups) $^{1,2}$,
 which are nonlinear
generalizations of the usual Lie algebras to which they reduce for
appropriate values of the deformation parameter(s), have been finding
applications as the dynamical symmetry algebras of several physical
systems. For the boson realization of these algebras, various kinds
of deformed oscillators have been introduced $^{3-7}$
and unification schemes for them have been suggested (see $^8$ for
a list of references).

Furthermore, generalized deformed su(2) algebras have been introduced
$^{9,10}$, in a way that their representation theory remains as close
as possible to the usual su(2) one. It will be shown here that several
physical systems (two identical particles in two dimensions $^{11}$,
isotropic oscillator and Kepler system in a 2-dim curved space $^{12}$),
as well as the
quadratic algebra QH(3) $^{13}$ and the finite W algebra $\bar {\rm W}_0$
$^{14}$ can be accommodated within this scheme. The advantage this
unification offers is that the representation theory of the generalized
deformed su(2) algebras is known $^9$.

In addition, generalized parafermionic oscillators have been introduced
$^{15}$, in analogy to generalized deformed oscillators $^{16}$.
It will be shown here that several physical systems (isotropic oscillator
and Kepler system in a 2-dim curved space $^{12}$, Fokas--Lagerstrom
potential
$^{17}$, Smorodinsky--Winternitz potential $^{18}$, Holt potential
$^{19}$) and mathematical constructions (generalized deformed su(2)
algebra $^{9,10}$, finite W algebras $\bar {\rm W}_0$ $^{14}$ and
W$_3^{(2)}$ $^{20-22}$) can be accommodated within this framework.
As a by-product the fact that the Holt potential is characterized by the
W$_3^{(2)}$ symmetry occurs.

 In section 2 the  cases related  to the generalized
deformed su(2) algebra will be studied, while in section 3 the systems
 related to generalized deformed
parafermionic oscillators will be  considered. Section 4 will contain
discussion of the present results and plans for further work.

\bigskip
{\bf 2. Generalized deformed su(2) algebras}
\medskip

Generalized deformed su(2) algebras having representation theory similar
to that of the usual su(2) have been constructed in $^9$. It has
been proved that it is possible to construct an algebra
$$ [J_0, J_{\pm}]=\pm J_{\pm}, \qquad [J_+,J_-]=\Phi(J_0(J_0+1))-
\Phi(J_0(J_0-1)),\eqno(1)$$
where $J_0$, $J_+$, $J_-$ are the generators of the algebra and
$\Phi(x)$ is any increasing entire function defined for $x\geq -1/4$.
Since this algebra is characterized by the function $\Phi$, we use for it
the symbol su$_{\Phi}$(2). The appropriate basis $|l,m>$ has the
properties
$$ J_0|l,m> = m |l,m>, \eqno(2)$$
$$ J_+|l,m> = \sqrt{\Phi(l(l+1))-\Phi(m(m+1))} |l,m+1>, \eqno(3)$$
$$ J_-|l,m> = \sqrt{\Phi(l(l+1))-\Phi(m(m-1))} |l, m-1>,\eqno(4) $$
where
$$ l=0, {1\over 2}, 1, {3\over 2}, 2, {5\over 2}, 3, \ldots,\eqno(5)$$
  and
$$ m= -l, -l+1, -l+2, \ldots, l-2, l-1, l.\eqno(6)$$
The Casimir operator is
$$ C= J_-J_+ +\Phi(J_0(J_0+1))=J_+J_-+\Phi(J_0(J_0-1)),
\eqno(7)$$
its eigenvalues indicated by
$$C |l,m> = \Phi(l(l+1)) |l,m>.\eqno(8)$$
The usual su(2) algebra is recovered for
$$ \Phi(x(x+1))= x(x+1),\eqno(9)$$
while the quantum algebra su$_q$(2)
$$ [J_0, J_{\pm}]=\pm J_{\pm}, \qquad [J_+, J_-]=[2 J_0]_q,\eqno(10) $$
occurs for
$$ \Phi(x(x+1))= [x]_q [x+1]_q,\eqno(11)$$
with $q$-numbers defined as
$$ [x]_q ={q^x-q^{-x}\over q-q^{-1}}.\eqno(12)$$

The su$_{\Phi}$(2) algebra occurs in several cases, in which the rhs of the
last equation in (1) is an odd function of $J_0$.

\medskip
{\it 2.1 Two identical particles in two dimensions}
\medskip

Let us consider the system of two identical particles in two
dimensions. For identical particles observables of the system have to be
invariant under exchange of particle indices.
A set of appropriate observables in this case is $^{11}$
$$ u=(x_1)^2+(x_2)^2, \qquad v=(x_1)^2-(x_2)^2, \qquad w=2 x_1 x_2,\eqno(13)$$
$$ U=(p_1)^2+(p_2)^2, \qquad V=(p_1)^2-(p_2)^2, \qquad W=2p_1 p_2,\eqno(14)$$
$$ C_1= {1\over 4} (x_1 p_1+p_1 x_1), \qquad C_2={1\over 4} (x_2 p_2+
p_2 x_2), \qquad M= x_1 p_2+x_2 p_1,\eqno(15)$$
where the indices 1 and 2 indicate the two particles.
These observables are known to close an sp(4,R) algebra. A representation
of this algebra can be constructed $^{11,23}$ using one arbitrary constant
$\eta$ and three matrices $Q$, $R$, and $S$ satisfying the commutation
relations
$$ [S,Q]= -2iR, \qquad [S,R]=2iQ, \qquad [Q,R]= -8iS (\eta-2 S^2).\eqno(16)$$
The explicit expressions of the generators of sp(4,R) in terms of $\eta$,
$S$, $Q$, $R$ are given in $^{11}$ and need not be repeated here.
Defining the operators
$$ X=Q-iR, \qquad Y=Q+iR, \qquad S_0= {S\over 2},\eqno(17)$$
one can see that the commutators of eq. (16) take the form
$$ [S_0,X]=X, \qquad [S_0,Y]=-Y, \qquad [X,Y]=32S_0 (\eta-8 (S_0)^2),
\eqno(18)$$
which is a deformed version of su(2).
It is clear that the algebra of eq. (18) is a special case of an
su$_{\Phi}$(2) algebra with structure function
$$ \Phi(J_0(J_0+1))= 16 \eta J_0(J_0+1) -64 (J_0(J_0+1))^2.\eqno(19)$$
The condition that $\Phi(x)$ has to be an increasing function of $x$
implies the restriction $x<\eta/8$.

\medskip
{\it 2.2 Kepler problem in 2-dim curved space}
\medskip

Studying the Kepler problem in a two-dimensional curved space with
constant curvature
$\lambda$ one finds the algebra $^{12}$
$$ [L, R_{\pm}]=\pm R_{\pm}, \qquad  [R_-, R_+]= F\left(L+{1\over 2}\right)-
F\left(L-{1\over 2}\right), \eqno(20)$$
where $$F(L)= \mu^2 + 2H L^2 -\lambda L^2\left(L^2-{1\over 4}\right).
\eqno(21)$$
It is then easy to see that
$$[R_+, R_-]= 2L\left(-2H+{\lambda\over 4}\right) + 4\lambda L^3,\eqno(22)$$
which corresponds to an su$_{\Phi}$(2) algebra with
$$ \Phi(J_0(J_0+1)) = \left(-2H+{\lambda\over 4}\right) J_0(J_0+1) +\lambda
(J_0(J_0+1))^2.\eqno(23)$$
For $\Phi(x)$ to be an increasing function, the condition
$$ \lambda x > H -{\lambda\over 8}\eqno(24)$$
has to be obeyed.

{\it 2.3 Isotropic oscillator in 2-dim curved space}

 In the case of the isotropic oscillator in a two-dimensional curved
space with constant curvature $\lambda$ one finds the algebra $^{12}$
$$ [L,S_{\pm}] =\pm 2 S_{\pm}, \qquad [S_-, S_+]= G(L+1)-G(L-1),\eqno(25)$$
with $$G(L)= H^2-\left(\omega^2+{\lambda^2\over 4}+\lambda H\right) L^2
+{1\over 4} \lambda^2 L^4.\eqno(26)$$
Using $\tilde L = L/2$ one easily sees that
$$ [\tilde L, S_{\pm}]= \pm S_{\pm}, \qquad
[S_+, S_-]= 8\tilde L \left(\omega^2-{\lambda^2\over 4}+\lambda H\right)
-16 \lambda^2 \tilde L^3,\eqno(27)$$
which corresponds to an su$_{\Phi}$(2) algebra with
$$ \Phi(J_0(J_0+1))= 4\left(\omega^2-{\lambda^2\over 4}+\lambda H\right)
J_0(J_0+1)-4\lambda^2  (J_0(J_0+1))^2.\eqno(28)$$
For $\Phi(x)$ to be an increasing function, the condition
$$ x< {1\over 2 \lambda^2} \left( \omega^2 -{\lambda^2\over 4}
+\lambda H \right)\eqno(29) $$
has to be satisfied.

\medskip
{\it 2.4 The quadratic Hahn algebra QH(3)}
\medskip

 The quadratic Hahn algebra QH(3) $^{13}$
$$ [K_1, K_2]= K_3, \eqno(30)$$
$$ [K_2, K_3]= A_2 K_2^2 + C_1 K_1 + D K_2 + G_1,\eqno(31)$$
$$ [K_3, K_1]= A_2(K_1 K_2+K_2 K_1) + C_2 K_2 +D K_1 +G_2,\eqno(32)$$
can be put in correspondence to an su$_{\Phi}$(2) algebra in the special
case in which $C_1=-1$ and $D=G_2=0$. The equivalence can be seen $^{24}$
by defining the operators
$$ J_1={J_++J_-\over 2}, \qquad J_2={J_+-J_-\over 2i}.\eqno(33)$$
Then the su$_{\Phi}$(2) commutation relations can be written as
$$ [J_0, J_1]= iJ_2, \qquad [J_0, J_2]=-i J_1,\eqno(34)$$
$$[J_1, J_2]= {i\over 2} (\Phi(J_0(J_0+1))-\Phi(J_0(J_0-1)).\eqno(35)$$
Subsequently one can see that the two algebras are equivalent for
$$ K_1= J_1 + A_2 J_0^2 + G_1, \qquad K_2=J_0, \qquad K_3= -i J_2,\eqno(36)$$
and
$$ \Phi(J_0(J_0+1))= -(2 A_2 G_1 + C_2) J_0(J_0+1) -A_2^2 (J_0(J_0+1))^2.
\eqno(37)$$
For $\Phi(x)$ to be an increasing function, the condition
$$ x< -{2 A_2 G_1 + C_2 \over 2 A_2^2} \eqno(38)$$
has to be obeyed.

\medskip
{\it 2.5 The finite W algebra $\bar{\rm W}_0$}
\medskip

 It is worth remarking that the finite W algebra $\bar{\rm W}_0$ $^{14}$
$$[U_0, L_0^{\pm}]= \pm L_0^{\pm}, \eqno(39)$$
$$ [L_0^+, L_0^-]= (-k(k-1)-2(k+1)h) U_0 +2 (U_0)^3,\eqno(40)$$
is also an su$_{\Phi}$(2) algebra with
$$ \Phi(J_0(J_0+1))= \left(-{k(k-1)\over 2}-(k+1)h\right) J_0(J_0+1) +
{1\over 2}  (J_0(J_0+1))^2.\eqno(41)$$
For $\Phi(x)$ to be an increasing function, the condition
$$ x > {k(k-1)\over 2} +(k+1)h \eqno(42)$$
has to be satisfied.

\bigskip
In all of the above cases the representation theory of the su$_{\Phi}$(2)
algebra immediately follows from eqs. (2)--(4). In each case the range of
values of the free parameters is limited by the condition that
$\Phi(x)$ has to be an increasing entire function defined for $x\geq -1/4$.
The results of this section are summarized in Table 1.

\bigskip
{\bf 3. Generalized deformed parafermionic oscillators}
\medskip

The relation of the above mentioned algebras, and of additional ones,
to generalized deformed parafermions is also worth studying.

A deformed oscillator $^{8,16}$
can be defined by the algebra generated by the
operators $\big\{ 1,a,a^+,N\big\}$ and the {\it structure
function} $\Phi (x)$, satisfying the relations:
$$\left[ a , N \right] = a, \quad  \quad
 \left[ a^+ , N \right] = -a^+ , \eqno(43)$$
and
$$ a^+a=\Phi(N)=[N], \qquad aa^+=\Phi(N+1)=[N+1], \eqno(44)$$
where $\Phi(x)$ is a positive analytic function with
$\Phi(0)=0$ and $N$ is the number operator.
{}From eq. (44)
 we conclude that:
$$N= \Phi^{-1}\left( a^+ a \right), \eqno(45)$$
and that the following commutation and
anticommutation relations are obviously satisfied:
$$ \left[ a,a^+ \right] = [N+1]-[N], \qquad
   \left\{ a,a^+ \right\} = [N+1]+[N] .\eqno(46)$$

The {\it structure  function} $\Phi(x)$ is characteristic to the deformation
scheme.  In Table 2 (cases i--iv)
 the structure functions corresponding to various deformed
oscillators are given.

The generalized deformed algebras possess a Fock space of
 eigenvectors
$|0>$, $|1>$, $\ldots$, $|n>$, $\ldots$
of the number operator $N$
$$N|n>=n|n>,\quad <n|m>=\delta_{nm}, \eqno(47) $$
if the {\it vacuum state} $|0>$ satisfies the following relation:
$$ a|0>=0. \eqno(48)$$
 These eigenvectors are generated
 by the formula:
 $$ \vert n >= {1 \over \sqrt{ [n]!}} {\left( a^+ \right)}^n \vert 0 >,
\eqno(49) $$
where
 $$[n]!=\prod_{k=1}^n [k]= \prod_{k=1}^n \Phi(k). \eqno(50) $$
The generators  $a^+$ and $a$ are the creation and
destruction operators of this deformed oscillator algebra:
$$a\vert n> = \sqrt{[n]} \vert n-1>,\qquad
 a^+\vert n> = \sqrt{[n+1]} \vert n+1>. \eqno(51) $$

 It has been proved $^{15}$ that any generalized
deformed parafermionic algebra of order $p$ can be written as a generalized
oscillator with structure function
$$ F(x)= x (p+1-x) (\lambda +\mu x+\nu x^2 +\rho x^3 +\sigma x^4 +\ldots),
\eqno(52)$$
where $\lambda$, $\mu$, $\nu$, $\rho$, $\sigma$, \dots are real constants
satisfying the conditions
$$ \lambda + \mu x + \nu x^2 + \rho x^3 + \sigma x^4 +\ldots > 0, \qquad
x \in \{ 1,2,\ldots, p\}.\eqno(53)$$

\medskip
{\it 3.1 The su$_{\Phi}$(2) algebra}
\medskip

Considering an su$_{\Phi}$(2) algebra $^9$ with structure function
$$ \Phi(J_0(J_0+1))= A J_0(J_0+1) + B (J_0(J_0+1))^2 + C (J_0(J_0+1))^3,
\eqno(54) $$
and making the correspondence
$$ J_+ \to A^{\dag}, \qquad J_-\to A, \qquad J_0\to N,\eqno(55)$$
one finds by equating the rhs of the first of eq. (46) and the last of eq.
(1) that the su$_{\Phi}$(2) algebra is equivalent to a generalized
deformed parafermionic  oscillator of the form
$$F(N)= N (p+1-N)$$ $$ [ -(p^2(p+1)C +p B)+ (p^3 C +(p-1)B) N $$ $$+
((p^2-p+1)C +B) N^2+ (p-2) C N^3 + C N^4], \eqno(56)$$
if the condition
$$ A+ p(p+1) B + p^2 (p+1)^2 C =0 \eqno(57)$$
holds. The condition of eq. (53) is always satisfied for $B>0$ and $C>0$.

In the special case of $C=0$ one finds that the su$_{\Phi}$(2) algebra
with structure function
$$ \Phi(J_0(J_0+1))= A J_0(J_0+1) + B (J_0(J_0+1))^2\eqno(58)$$
is equivalent to a generalized deformed parafermionic oscillator
characterized by
$$ F(N)= B N (p+1-N) (-p+(p-1)N+ N^2),\eqno(59)$$
if the condition
$$ A+ p(p+1) B=0\eqno(60)$$
is satisfied. The condition of eq. (53) is satisfied for $B>0$.

Including higher powers of $J_0(J_0+1)$ in eq. (54) results in higher powers
of $N$ in eq. (56) and higher powers of $p(p+1)$ in eq. (57). If, however,
one sets $B=0$ in eq. (58), then eq. (59) vanishes, indicating that no
parafermionic oscillator equivalent to the usual su(2) rotator can be
constucted.

\medskip
{\it 3.2 The finite W algebra $\bar {\rm W}_0$}
\medskip

 The $\bar {\rm W}_0$ algebra $^{14}$
of eqs (39)-(40) is equivalent to a generalized
deformed parafermionic algebra with
$$F(N)= N (p+1-N) {1\over 2} (-p+(p-1)N+N^2),\eqno(61)$$
provided that the condition
$$ k(k-1)+2(k+1)h=p(p+1) \eqno(62)$$
holds. One can easily check that the condition of eq. (53) is satisfied
without any further restriction.

\medskip
{\it 3.3 Isotropic harmonic oscillator in a 2-dim curved space}

The algebra of the isotropic harmonic oscillator in a 2-dim curved space with
constant curvature $\lambda$ for finite representations can be put in the
form $^{25}$
$$F(N)= 4 N (p+1-N) \left(\lambda (p+1-N)+\sqrt{\omega^2+\lambda^2/4}\right)
\left( \lambda N +\sqrt{\omega^2+\lambda^2/4}\right),\eqno(63)$$
the relevant energy eigenvalues being
$$ E_p = \sqrt{\omega^2 +{\lambda^2\over 4}} (p+1) +{\lambda\over 2}
(p+1)^2,\eqno(64)$$
where $\omega$ is the angular frequency of the oscillator. It is clear
that the condition of eq. (53) is satisfied without any further
restrictions.

\medskip
{\it 3.4 The Kepler problem in a 2-dim curved space}

The algebra of the Kepler problem in a 2-dim curved space with constant
curvature
$\lambda$ for finite representations can be put in the form $^{25}$
$$ F(N)= N(p+1-N) \left( {4\mu^2\over (p+1)^2} +
\lambda {(p+1-2N)^2\over 4}\right),\eqno(65)$$
the corresponding energy eigenvalues being
$$ E_p = -{2\mu^2\over (p+1)^2}+\lambda {p(p+2)\over 8},\eqno(66)$$
where $\mu$ is the coefficient of the $-1/r$ term in the Hamiltonian. It is
clear that the restrictions of eq. (53) are satisfied automatically.

\medskip
{\it 3.5 The Fokas--Lagerstrom potential}

The Fokas--Lagerstrom potential $^{17}$ is described by the Hamiltonian
$$ H= {1\over 2} (p_x^2+p_y^2)+{x^2\over 2} + {y^2\over 18}.\eqno(67)$$
It is therefore an anisotropic oscillator with ratio of frequencies 3:1.
For finite representations it can be seen $^{25}$ that the relevant
algebra can be put in the form
$$ F(N)=16 N (p+1-N) \left(p+{2\over 3}-N\right) \left( p+{4\over 3}-N\right)
\eqno(68)$$
for energy eigenvalues $E_p=p+1$, or in the form
$$ F(N)= 16 N (p+1-N) \left( p+{2\over3}-N\right) \left( p+{1\over3}-N\right)
\eqno(69)$$
for eigenvalues $E_p=p+2/3$, or in the form
$$ F(N)=16 N (p+1-N) \left(p+{5\over 3}-N\right) \left( p+{4\over 3}-N\right)
\eqno(70)$$
for energies $E_p=p+4/3$. In all cases it is clear that the restrictions of
eq. (53) are satisfied.

\medskip{\it 3.6 The Smorodinsky--Winternitz potential}

The Smorodinsky--Winternitz potential $^{18}$ is described by the
Hamiltonian
$$ H={1\over 2} (p_x^2+p_y^2) +k (x^2+y^2) +{c\over x^2},\eqno(71)$$
i.e. it is a generalization of the isotropic harmonic oscillator in
two dimensions. For finite representations it can be seen $^{25}$ that the
relevant algebra takes the form
$$F(N)= 1024 k^2 N (p+1-N) \left( N+{1\over2}\right) \left( p+1+{\sqrt{1+
8 c}\over 2} -N\right)\eqno(72)$$
for $c\geq -1/8$ and energy eigenvalues
$$ E_p= \sqrt{8k}\left(p+{5\over 4}+{\sqrt{1+8c}\over 4}\right), \qquad
p=1,2,\ldots.\eqno(73)$$
In the special case of $-1/8 \leq c \leq 3/8$ and energy eigenvalues
$$ E_p= \sqrt{8k} \left( p+{5\over 4}-{\sqrt{1+8c}\over 4}\right), \qquad
p=1, 2, \ldots\eqno(74)$$
the relevant algebra is
$$F(N)=1024 k^2 N (p+1-N)\left( N+{1\over 2}\right) \left( p+1-
{\sqrt{1+8c}\over 2}-N\right).\eqno(75)$$
In both cases the restrictions of eq. (53) are satisfied.

\medskip
{\it 3.7 Two identical particles in two dimensions}
\medskip

Using the same procedure as above, the algebra of eq. (18)  can be put
in correspondence with a parafermionic oscillator characterized by
$$ F(N)= N(p+1-N) 64 (p+(1-p)N -N^2),\eqno(76)$$
if the condition
$$ \eta = 4p(p+1)\eqno(77)$$
holds. However, the condition of eq. (53) is violated in this case.

\medskip
{\it 3.8 The quadratic Hahn algebra QH(3)}
\medskip

For the quadratic Hahn algebra QH(3) of eqs (30)-(32) one obtains the
parafermionic oscillator with
$$ F(N)= N (p+1-N) A_2^2 (p+(1-p)N-N^2),\eqno(78)$$
if the condition
$$ p(p+1) A_2^2+ 2 A_2 G_1 + C_2 =0\eqno(79)$$
holds. Again, eq. (53) is violated in this case.

\medskip
{\it 3.9 The finite W algebra W$_3^{(2)}$}
\medskip

The finite W algebra W$_3^{(2)}$ $^{20-22}$
is characterized by the commutation relations
$$ [H,E]=2E, \qquad [H,F]=-2F, \qquad [E,F]=H^2+C, \eqno(80)$$
$$ [C,E]=[C,F]=[C,H]=0.\eqno(81)$$
Defining $\tilde H=H/2$ these can be put in the form
$$ [\tilde H, E]=E, \qquad [\tilde H, F]=-F, \qquad [E,F]=4 \tilde H^2 +C,
\eqno(82)$$
$$ [C,E]=[C,F]=[C,\tilde H]=0.\eqno(83)$$
This algebra is equivalent to a parafermionic oscillator with
$$ F(N)={2\over 3} N(p+1-N) ( 2 p-1+2 N),\eqno(84)$$
provided that the condition
$$ C=-{2\over 3} p (2p+1)\eqno(85)$$
holds. One can easily see that the condition of eq. (53) is satisfied
without any further restriction.

\medskip \vfill\eject
{\it 3.10 The Holt potential}
\medskip

The Holt potential $^{19}$
$$ H = {1\over 2} (p_x^2+p_y^2) +(x^2+4y^2)+{\delta \over x^2}\eqno(86) $$
is a generalization of the harmonic oscillator potential with a ratio of
frequencies 2:1. The relevant algebra can be put $^{25}$ in the form
of an oscillator with
$$ F(N)= 2^{23/2}  N (p+1-N) \left( p+1+{\sqrt{1+8\delta}\over 2} -N\right),
\eqno(87)$$
where $(1+8\delta)\geq 0$, the relevant energies being given by
$$ E_p= \sqrt{8} \left(p+1+{\sqrt{1+8\delta}\over 4}\right).\eqno(88)$$
In this case it is clear that the condition of
eq. (53) is always satisfied without any further restrictions.

In the special case $-{1\over 8} \leq \delta \leq {3\over 8}$ one
obtains $^{25}$
$$ F(N)= 2^{23/2} N (p+1-N) \left(p+1-{\sqrt{1+8\delta}\over 2} -N\right),
\eqno(89)$$
the relevant energies being
$$ E_p=\sqrt{8} \left( p+1-{\sqrt{1+8\delta}\over 4}\right).\eqno(90)$$
The condition of eq. (53) is again satisfied without any further restrictions
within the given range of $\delta$ values.

The deformed oscillator commutation relations in these cases take the
form
$$ [N,A^{\dagger}]=A^{\dagger}, \qquad [N,A]=-A,\eqno(91)$$
$$ [A, A^{\dagger}]= 2^{23/2} \left( 3 N^2 -N\left( 4p+1 \pm \sqrt{1+8\delta}
\right) +p^2 \pm {1\over 2} p \sqrt{1+8\delta}\right).\eqno(92)$$
It can easily be seen that they are the same as the W$_3^{(2)}$
commutation relations $^{20-22}$ with the identifications
$$ F=\sigma A^{\dagger}, \qquad E=\rho A,\qquad C=f(p), \qquad
H=-2N+k(p),\eqno(93)$$
where
$$ \rho\sigma = 2^{-19/2} /3,\qquad k(p)={1\over 3} \left( 4p+1\pm
\sqrt{1+8\delta}\right),\eqno(94)$$
$$ f(p)= {2\over 9} \left( 14 p^2 +4 p\pm (7p+1)\sqrt{1+8\delta}+1+4\delta
\right).\eqno(95)$$
It is thus shown that the Holt potential possesses the W$_3^{(2)}$
symmetry.

The results of this section are summarized in Table 2.

{\bf 4. Discussion}

In conclusion, we have shown that several physical systems (two identical
particles in two dimensions, isotropic oscillator and Kepler system in a
2-dim curved space) and mathematical structures (quadratic Hahn algebra
QH(3), finite W algebra $\bar {\rm W}_0$) are identified with a
generalized deformed su(2) algebra, the representation theory of which
is known. The results are summarized in Table 1.
Furthermore, the generalized deformed parafermionic oscillator
is found to describe several physical systems (isotropic oscillator and
Kepler system in a curved space, Fokas--Lagerstrom,
Smorodinsky--Winternitz and Holt potentials)
and mathematical constructions (generalized deformed su(2) algebras,
finite W algebras $\bar {\rm W}_0$ and W$_3^{(2)}$).
The results are summarized in Table 2.
The framework of the
generalized deformed parafermionic oscillator is more general than the
generalized deformed su(2) algebra, since in the rhs of the relevant
basic commutation relation in the former case (first equation in eq. 46)
both odd and even
powers are allowed, while in the latter case (eq. 1) only odd powers are
allowed.

The relevance of deformed oscillator algebras, finite W algebras and
qua\-dra\-tic algebras in the study of the symmetries of the anisotropic
quantum harmonic oscillator in two and three dimensions is receiving
attention.

{\bf Acknowledgments}

Support by CEC under contract ERBCHBGCT930467 is gratefully acknowledged
by one of the authors (DB).

\vfill\eject

\parindent=0pt

\centerline{\bf References}

\ref {1.} V. G. Drinfeld, in {\it Proceedings of the International Congress of
Mathematicians}, ed. A. M. Gleason (American Mathematical Society,
Providence, RI, 1986) p. 798.

\ref {2.}
M. Jimbo, {\it Lett. Math. Phys.}  {\bf 11}, 247 (1986).

\ref {3.}
M. Arik and D. D. Coon, {\it J. Math. Phys.} {\bf  17}, 524 (1976).

\ref {4.}
V. V. Kuryshkin, {\it Annales de la Fondation Louis de Broglie} {\bf  5},
111 (1980).

\ref {5.}
L. C. Biedenharn, {\it J. Phys. A} {\bf 22}, L873 (1989).

\ref {6.}
A. J. Macfarlane, {\it J. Phys. A} {\bf 22}, 4581 (1989).

\ref {7.}
C. P. Sun and H. C. Fu, {\it J. Phys. A} {\bf  22}, L983 (1989).

\ref {8.}
D. Bonatsos and C. Daskaloyannis, {\it Phys. Lett. B} {\bf 307}, 100 (1993).

\ref {9.}
D. Bonatsos, C. Daskaloyannis and P. Kolokotronis, {\it J. Phys. A} {\bf 26},
L871 (1993).

\ref {10.}
F. Pan, {\it J. Math. Phys.} {\bf 35}, 5065 (1994).

\ref {11.}
J. M. Leinaas and J. Myrheim, {\it Int. J. Mod. Phys. A} {\bf 8}, 3649 (1993).

\ref {12.}
P. W. Higgs, {\it J. Phys. A} {\bf 12}, 309 (1979).

\ref {13.}
Ya. I. Granovskii, I. M. Lutzenko and A. S. Zhedanov, {\it Ann. Phys.}
 {\bf 217}, 1 (1992).

\ref {14.}
P. Bowcock, U. Durham preprint DTP 94-5 (1994).

\ref {15.}
C. Quesne, {\it Phys. Lett. A} {\bf  193}, 245 (1994).

\ref {16.}
C. Daskaloyannis, {\it J. Phys. A} {\bf 24}, L789 (1991).

\ref {17.}
A. S. Fokas and P. A. Lagerstrom, {\it J. Math. Anal. Appl.} {\bf 74}, 325
 (1980).

\ref {18.}
P. Winternitz, Ya. A. Smorodinsky, M. Uhlir and I. Fris, {\it Yad. Fiz.}
{\bf 4}, 625
(1966) [{\it Sov. J. Nucl. Phys.} {\bf 4}, 444 (1966)].

\ref {19.}
C. R. Holt, {\it J. Math. Phys.} {\bf 23}, 1037 (1982).

\ref {20.}
T. Tjin, {\it Phys. Lett. B} {\bf 292}, 60 (1992).

\ref {21.}
J. de Boer and T. Tjin, {\it Commun. Math. Phys.} {\bf 158}, 485 (1993).

\ref {22.}
T. Tjin, Ph.D. thesis, U. Amsterdam (1993).

\ref {23.}
F. Barbarin, E. Ragoucy and P. Sorba, {\it Nucl. Phys. B} {\bf 442}, 425
 (1995).

\ref {24.}
A. S. Zhedanov, {\it Mod. Phys. Lett. A} {\bf  7}, 507 (1992).

\ref {25.}
D. Bonatsos, C. Daskaloyannis and K. Kokkotas, {\it Phys. Rev. A} {\bf  50},
3700 (1994).

\ref {26.}
Y. Ohnuki and S. Kamefuchi, {\it Quantum field theory and parastatistics}
(Springer, Berlin, 1982).

\ref {27.}
K. Odaka, T. Kishi and S. Kamefuchi, {\it J. Phys. A} {\bf  24}, L591 (1991).

\newpage

\begin{table}[bth]
\begin{center}
\caption{ Structure functions of generalized deformed su(2) algebras.
For conditions of validity for each of them see the corresponding
subsection of the text. }
\bigskip\bigskip
\begin{tabular}{|c c p{2.0 in}|}
\hline
\ & $\Phi(J_0(J_0+1))$ & Reference \\
\hline\hline
\romannumeral 1 & $J_0(J_0+1)$ & usual su(2) \\[0.05in]
\romannumeral 2 & $ [J_0]_q [J_0+1]_q $ & su$_q$(2) $^{5-7}$
\\[0.05in]
\romannumeral 3 & $ 16 \eta J_0(J_0+1) -64 (J_0(J_0+1))^2$ &
           2 identical particles in 2-dim $^{11}$\\[0.05in]
\romannumeral 4 & $ \left(-2H+{\lambda\over 4}\right) J_0(J_0+1) +\lambda
(J_0(J_0+1))^2$ & Kepler system in 2-dim curved space $^{12}$\\[0.05in]
\romannumeral 5 & $ 4\left( \omega^2-{\lambda^2\over 4}+\lambda H\right)
J_0(J_0+1)-4 \lambda^2 (J_0(J_0+1))^2 $ & isotropic oscillator in
2-dim curved space $^{12}$ \\[0.05in]
\romannumeral 6 & $ -(2 A_2 G_1 + C_2) J_0(J_0+1) -A_2^2 (J_0(J_0+1))^2 $ &
quadratic Hahn algebra QH(3) $^{13}$\\[0.05in]
\romannumeral 7 & $ \left(-{k(k-1)\over 2}-(k+1)h\right) J_0(J_0+1) +
{1\over 2}  (J_0(J_0+1))^2 $ & finite W algebra $\bar {\rm W}_0$
$^{14}$\\[0.05in]
\hline
\end{tabular}
\end{center}
\end{table}

\begin{table}[bth]
\begin{center}
\caption{ Structure functions of deformed oscillators. For conditions
of validity and further explanations in the case of the various
generalized deformed parafermionic oscillators see the corresponding
subsection in the text. }
\bigskip\bigskip
\begin{tabular}{|c c p{2.0 in}|}
\hline
\ & $F(N)$ & Reference \\
\hline\hline
\romannumeral 1 & $N$ & harmonic  oscillator \\[0.05in]
\romannumeral 2 &  ${ {q^N- q^{-N} }  \over {q- q^{-1} } }= [N]_q $ &
$q$-deformed harmonic oscillator $^{5-7}$\\[0.05in]
\romannumeral 3 & $N(p+1-N)$ & parafermionic oscillator $^{26}$
 \\[0.05in]
\romannumeral 4 & $ [N]_q [p+1-N]_q$ & $q$-deformed parafermionic oscillator
$^{27}$ \\[0.05in]
\romannumeral 5 & $ N (p+1-N) (\lambda +\mu N+\nu N^2 +\rho N^3 +\sigma N^4
+\ldots)$ & generalized deformed parafermionic oscillator $^{15}$ \\[0.05in]
\romannumeral 6 & $ N (p+1-N) [ -(p^2(p+1)C +p B)+ (p^3 C +(p-1)B) N $ &
3-term su$_{\Phi}$(2) algebra \\
          &$+((p^2-p+1)C +B) N^2+ (p-2) C N^3 + C N^4]$ & (eq. 54)  \\[0.05in]
\romannumeral 7 & $ B N (p+1-N) (-p+(p-1)N+ N^2)$ & 2-term su$_{\Phi}$(2)
algebra (eq.~58)\\[0.05in]
\romannumeral 8 & $ N (p+1-N) {1\over 2} (-p+(p-1)N+N^2)$ & finite W algebra
$\bar {\rm W}_0$ $^{14}$ \\[0.05in]
\romannumeral 9 & $ 4 N (p+1-N) \left(\lambda (p+1-N)+\sqrt{\omega^2+
\lambda^2/4}\right) $ & isotropic oscillator in 2-dim \\
   & $ \left( \lambda N +\sqrt{\omega^2+\lambda^2/4}\right)$ &
curved space $^{12,25}$ \\[0.05in]
\romannumeral 10 & $ N(p+1-N) \left( {4\mu^2\over (p+1)^2} +
\lambda {(p+1-2N)^2\over 4}\right)$ & Kepler system in 2-dim curved space
$^{12,25}$\\[0.05in]
\romannumeral 11 & $16 N (p+1-N) \left(p+{2\over 3}-N\right) \left( p
+{4\over 3}-N\right)$ & Fokas--Lagerstrom potential \\
             & or  $ 16 N (p+1-N) \left( p+{2\over3}-N\right)
\left( p+{1\over3}-N\right)$ &  $^{17,25}$ \\
             & or $ 16 N (p+1-N) \left(p+{5\over 3}-N\right)
\left( p+{4\over 3}-N\right)$ & \\[0.05in]
\romannumeral 12 & $ 1024 k^2 N (p+1-N) \left( N+{1\over2}\right) \left(
p+1\pm {\sqrt{1+
8 c}\over 2} -N\right)$ & Smorodinsky-Winternitz potential $^{18,25}$
\\[0.05in]
\romannumeral 13 & ${2\over 3} N(p+1-N) ( 2 p-1+2 N)$ & finite W algebra
W$_3^{(2)}$ $^{20-22}$ \\[0.05in]
\romannumeral 14 & $ 2^{23/2}  N (p+1-N) \left( p+1\pm
{\sqrt{1+8\delta}\over 2} -N\right)$ & Holt potential $^{19,25}$
\\[0.05in]
\hline
\end{tabular}
\end{center}
\end{table}

\end{document}